\documentclass[journal=jpclcd,manuscript=letter,layout=twocolumn]{achemso}
\usepackage{mathtools}
\usepackage{units}
\usepackage{amssymb}
\usepackage{amsmath}
\usepackage{amsthm}
\usepackage{graphicx} 
\usepackage{subcaption}
\usepackage[usenames,dvipsnames,svgnames,table]{xcolor}

\newcommand{\blue}[1]{\textcolor{black}{#1}}

\newcommand{\suzuki}{Suzuki et.\ al.\ }
\newcommand{\michaud}{Michaud et.\ al.\ }

\title{Reply to ``Comment on `An Alternative Approach for the Determination of Mean Free Paths of 
       Electron Scattering in Liquid Water Based on Experimental Data{'}''}

\author{Axel Schild}
\affiliation{ETH Z\"urich,  Laboratorium f\"ur Physikalische Chemie,  8093 Z\"urich, Switzerland}
\email{axel.schild@phys.chem.ethz.ch}

\author{Michael Peper}
\affiliation{ETH Z\"urich,  Laboratorium f\"ur Physikalische Chemie,  8093 Z\"urich, Switzerland}

\author{Conaill Perry}
\affiliation{ETH Z\"urich,  Laboratorium f\"ur Physikalische Chemie,  8093 Z\"urich, Switzerland}

\author{Dominik Rattenbacher}
\affiliation{ETH Z\"urich,  Laboratorium f\"ur Physikalische Chemie,  8093 Z\"urich, Switzerland}
\alsoaffiliation{Max Planck Institute for the Science of Light, Staudtstr. 2, 91058 Erlangen, Germany}

\author{Hans Jakob W\"orner}
\affiliation{ETH Z\"urich,  Laboratorium f\"ur Physikalische Chemie,  8093 Z\"urich, Switzerland}

\SectionsOff 

\begin{document}
  
  \maketitle
  
  \emph{Note: The comment which is reply is referring to was rejected for publication, hence this reply will also not be published.}
  \\
  
  In a recent comment,\cite{signorell2020comment} Ruth Signorell raises a number of issues that she considers to question the validity of our approach to determine mean free paths for electron scattering in liquid water\cite{schild2020} and our comparison with the results on amorphous ice by Michaud, Wen, and Sanche \cite{michaud2003}. Here, we show that these critiques are unjustified, being either unfounded or based on misconceptions by the author of the comment. We nevertheless welcome the opportunity to further clarify certain aspects of our work that we \blue{did} not discuss in detail in our letter\cite{schild2020}.
  Our reply is structured as the comment, i.e., the four main points of the comment are discussed individually.
  
  {\bf (1)}
  Signorell incorrectly claims that the effective attenuation length (EAL) as defined in our work is different from the definition used in the analysis of the measurements of \citet{suzuki2014}, which we take as input for our simulations. 
  
  \begin{figure}[htbp]
    \includegraphics[width=0.5\textwidth]{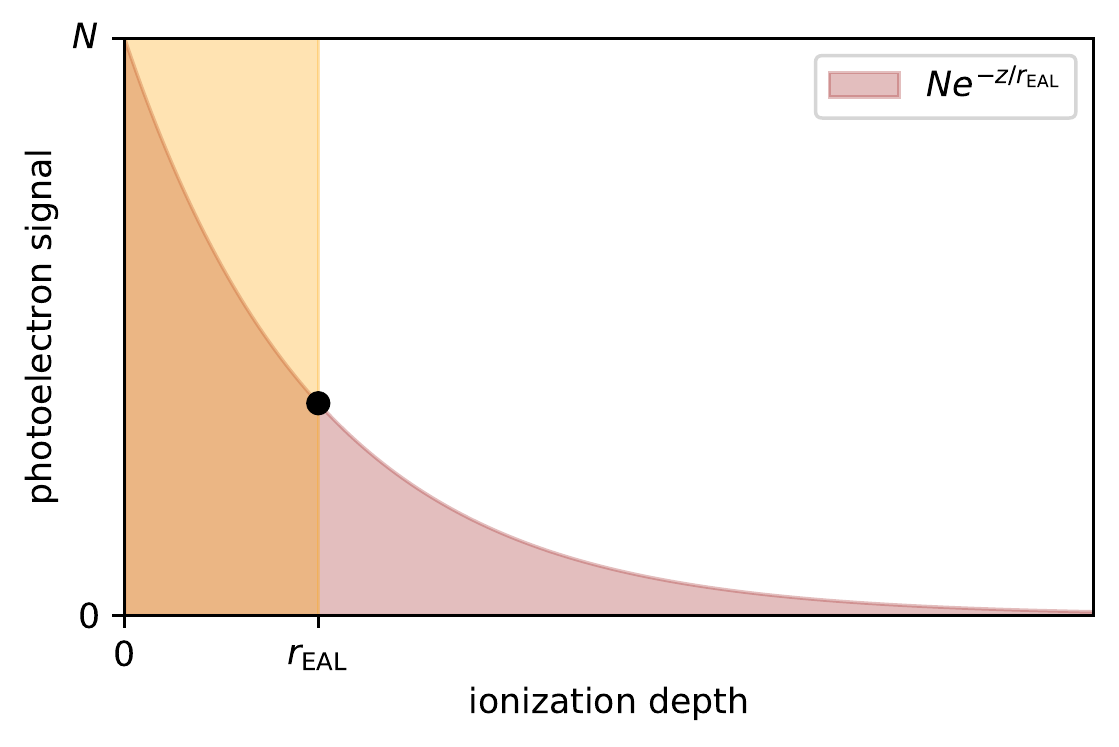}
    \caption{
      Exponential decay of the photoelectron signal with ionization depth.
      The area under the exponential function is equal to the area in the box given by $N \times r_{\rm EAL}$, which is used by \citet{suzuki2014} to determine $r_{\rm EAL}$ experimentally.
      }
    \label{fig:expo}
  \end{figure}
  
  Both our work and the analysis of \suzuki are based on the same standard definition of the EAL, i.e., the electron signal $S(z)$ detected outside the liquid decays exponentially with the distance from the point of ionization to the surface $z$,
  \begin{align}
    S(z) \propto \exp(-z/r_{\rm EAL}),
    \label{eq:eal}
  \end{align}
  and the EAL $r_{\rm EAL}$ is the width parameter of this distribution.
  \suzuki have applied this definition of the EAL, whereby they have realized that the total photoelectron signal integrated over an exponentially decaying depth profile is equal to the total photoelectron signal integrated over a constant depth profile of length $r_{\rm EAL}$,
  \begin{align}
    N \int\limits_0^{\infty} \exp(-z/r_{\rm EAL}) dz = N r_{\rm EAL},
  \end{align}
  graphically illustrated in Fig.\ \ref{fig:expo}.
  With this relation, it is clear that our definition of the EAL and the definition used by \suzuki are identical.
  
  At this point, we would like to mention that the decay of the electron signal as a function of $z$ is, indeed, exponential in our Monte-Carlo simulations, although elastic collisions are included. It follows that in our calculations the EAL and the mean-escape depth are equal\cite{jablonski1999}.
  We also note here that the probing depth of ca.\ \unit[6]{nm} reported by \citet{luckhaus2017} (see Fig. 4 in their article) at a photon energy of \unit[13.8]{eV} on the basis of their simulations is incompatible with the measured EAL of \unit[1.5-2]{nm} at the same photon energy by \citet{suzuki2014}.
  
  Signorell further remarks that the $\beta$-parameters from \citet{thurmer2013} are used in the analysis of the experimental data by \citet{suzuki2014}. This is correct, but it has nothing to do with the definition of the EAL. The $\beta$-parameters are only needed in the analysis of \suzuki because of the finite detection angle of their photoelectron spectrometer (see text preceding Eq.\ (5) of \citet{suzuki2014}), which results in a different detection efficiency of electrons originating from the gas and liquid phases.
  
  {\bf (2)}
  Signorell questions the validity of the cluster model by \blue{claiming (i) that} its convergence towards the liquid bulk is not demonstrated and \blue{(ii) that} differential scattering cross sections (DCS) are unsuitable for describing electron transport. 
    
Concerning (i), we note that our statement regarding the rapid convergence of the DCS for electron scattering with water clusters of increasing size is based on extensive calculations that will be published in a dedicated article in the near future. In the Supplemental Information (SI) of our letter\cite{schild2020}, we have exemplified this convergence for an electron kinetic energy (eKE) of \unit[50]{eV}, and for $($H$_2$O$)_n$ clusters with $n=3, 5, 7$.
The observed convergence of the DCS to a quasi-Gaussian shape of a characteristic width has also been observed at other electron kinetic energies. 
The changes of the DCS for $n>7$ are negligible for the purpose reported in our letter \cite{schild2020}.
   
  Concerning (ii), let us first discuss the assumptions underlying classical trajectory Monte-Carlo simulations (CTMCS) of electron transport (for a recent review, see e.g., \citet{nikjoo2016}). In such simulations, the complicated electron dynamics is approximated by classical trajectories that scatter at randomly chosen positions in a medium which is assumed to be homogeneous.
  Compared to a (totally infeasible) all-particle quantum dynamics simulation or a (feasible, if the corresponding time-dependent effective potential due to the other quantum particles could be obtained) one-electron quantum dynamics simulation, such a CTMCS involves many assumptions and approximations.
  One aim of our work was to improve CTMCS in a systematic and physically well-defined way by introducing a molecular-level description of electron scattering based on accurate quantum-scattering calculations.\cite{salvat2009}
  
  Prior to our work, elastic scattering cross sections from isolated water molecules\cite{pimblott1997} or the integral cross sections for amorphous ice from \michaud with a rescaling\cite{meesungnoen2002} have been used to describe collisions in CTMCS.
  Instead, we have used an {\it ab-initio} description of quantum scattering with the so-far best computationally tractable model of bulk liquid water.
  The difference between electron scattering with isolated molecules compared to liquid bulk is the presence of neighboring molecules in the latter case.
  The influence of these neighbors are mainly (a) that the electronic structure of the molecules and hence the scattering potential is modified due to solvation and (b) that multiple scattering between neighboring molecules lead to interference phenomena which are not contained in a classical trajectory description.
  Both of these effects are naturally included in our quantum-mechanical calculations of electron scattering with water clusters.
  As mentioned above, the remarkably rapid convergence of the DCS with increasing cluster size strongly suggests that relatively small water clusters are sufficient to successfully describe the dominant effects which distinguish electron scattering with isolated molecules compared to electron scattering in the liquid bulk.

  Concerning (ii), we also have to point out that there is a substantial misunderstanding in Signorell's comment that we have to rectify. For this purpose, we first recall the basic definitions of mean free paths, of transport cross sections (a.k.a.\ momentum-transfer cross sections), and of transport mean free paths,\cite{nikjoo2016} which Signorell actually refers to as ``mean free paths'' in her comment.\cite{signorell2020comment}
  The transport mean free path corresponds to a model in which the physical DCS is replaced by an isotropic DCS. This has the effect of combining multiple elastic collisions into a single model collision. As a consequence, the physical mean free path is replaced by the longer transport mean free path.
  The transport description is helpful for simulations of electron scattering in the bulk, as fewer collisions need to be simulated. However, it is unsuitable for the interpretation of measurements that are sensitive to the physical mean free paths. 
  An example of such a measurement technique that can be sensitive to the physical mean free path is attosecond interferometry \cite{rattenbacher2018}.

  The physical DCS and the associated mean-free paths, as used in our work, contain the information that is required to describe electron scattering within CTMCS on an event-by-event basis.\cite{plante2009,champion2012} This contrasts with transport mean-free paths and the associated isotropic DCS, which only reproduce the overall transport properties. Moreover, knowledge of the physical DCS is sufficient to completely determine the characteristic quantities in a transport description. Indeed, the DCS contains the information about the total scattering cross section
  \begin{align}
    \sigma = 2 \pi \int\limits_{0}^{\pi} \text{DCS}(\theta) \sin(\theta) {\mathrm d}\theta,
  \end{align}
  as well as the transport cross section 
  \begin{align}
    \sigma_{\rm mt} = 2 \pi \int\limits_{0}^{\pi} \text{DCS}(\theta) \left(1 - \cos(\theta)\right) \sin(\theta) {\mathrm d}\theta,
  \end{align}
  and thus the mean free path
  \begin{align}
    l_{\rm MFP} = \frac{1}{n \sigma} \label{eq:mfp_sigma}
  \end{align}
  and the transport mean free path
  \begin{align}
    l_{\rm MFP, tr} = \frac{1}{n \sigma_{\rm mt}}, \label{eq:mfp_sigma_mt}
  \end{align}
  where $n$ is the number density.
  It is important to correctly distinguish between (a) the mean free path obtained from Eq. (\ref{eq:mfp_sigma}), which we provide in table 1 of the SI of our letter ($l_{\rm ela}$) and which corresponds to the DCS describing scattering of an electron off our model cluster, (b) the mean free paths $r_{\rm MFP}$ for the liquid that we determine in the letter and that come from our simulations based on that DCS, and (c) the transport mean free path.
  We note that using the factors given in the SI, our mean free paths for the liquid may be converted to the transport mean free paths, if needed.
  Even when converting our mean-free paths to transport mean-free paths, they remain considerably shorter than the mean-free paths reported by \citet{michaud2003}.
  
  {\bf (3)}
  Seemingly due to a misconception regarding the difference between mean free paths and transport mean free paths as defined in \eqref{eq:mfp_sigma_mt}, Signorell questions the comparison of our results with results from amorphous ice obtained by \citet{michaud2003}.
  Both our mean free paths and those of \michaud are based on the same definition, i.e., the probability that a simulated electron trajectory has not scattered \blue{until} a distance $r$ from its origin or the previous collision is given by
  \begin{align}
   P(r) = \frac{e^{-r/r_{\rm MFP}}}{r_{\rm MFP}}
  \end{align}
  where $r_{\rm MFP}$ is the mean free path.
  It is important to note here that \citet{michaud2003} report ``integral cross sections'' throughout their work (see e.g.\ the abstract and table 2 of their article).
  They do not report \emph{transport} cross sections.
  This is further clarified in figure 3 of their article, where \michaud compare their cross sections with integral scattering cross sections of gas-phase water molecules.
  If the cross sections from \michaud were interpreted as transport cross sections, as done by Signorell, the comparison would have been to the transport cross sections of gas-phase water (see e.g.\ \citet{itikawa2005}).
  \michaud have further used these same integral scattering cross sections to calculate mean free paths (not transport mean free paths) in their table 3.
  We also note that we were not the first to point out that the inelastic mean free paths of \michaud, measured for amorphous ice, are much longer than independent results obtained for liquid water. This difference was previously mentioned by \citet{nikjoo2016}, \citet{shinotsuka2017} and \citet{nguyen2018}, among others. It is further worth pointing out that the scattering model used by \citet{michaud2003} does actually {\it not} assume isotropic elastic scattering into {\it all} spatial directions. \michaud instead use a simplified ``two-stream'' model, in which they assume that forward and backward elastic scattering are equally likely. Therefore, the results of \michaud cannot be translated into a three-dimensional CTMCS without further assumptions.
  
  Signorell also claims that the ``observed increase of scattering cross sections in clusters relative to bulk arises from the reduced dielectric shielding'' and refers to \citet{gartmann2018} to support this claim.\cite{signorell2020comment}
Careful inspection of the data shown in figure 4 of \citet{gartmann2018} actually reveals the opposite trend: The asymmetry parameters of small clusters are best described by the cross sections of \citet{michaud2003}, whereas those of larger clusters are increasingly better described by gas-phase cross sections. If ``dielectric shielding'' was the correct explanation for the observed effects, the trend should have been opposite, i.e., the $\beta$ parameters of large clusters should become increasingly better described by the condensed-phase scattering cross sections. We note that such a trend would also be required to explain why the electron-scattering cross sections (or mean-free paths) determined from water droplets by \citet{signorell2016} were indistinguishable from those of Michaud et al. \cite{michaud2003} within the quoted uncertainties. The trend observed by \citet{gartmann2018} (Fig. 4) is opposite and therefore inconsistent with ``dielectric shielding'' on one hand and the nearly perfect agreement of the electron scattering cross sections determined from droplets \cite{signorell2016} and amorphous ice \cite{michaud2003} on the other.
  
  {\bf (4)}
  Signorell claims that the uncertainties of the photoelectron angular distribution measurements by \citet{thurmer2013} would translate to large uncertainties of our determined values for the inelastic mean free path (a ``factor of two'' at \unit[20]{eV} and ``an order of magnitude'' at \unit[10]{eV}).\cite{signorell2020comment}
  
  Instead of showing the uncertainties in our letter, we opted for showing the sensitivity of our simulations on the input parameters in figure S3 of the SI. Based on that figure, it is possible to estimate how different input values for the $\beta$ parameter of the photoelectron angular distribution and the EAL change the resulting mean free paths, even without making a simulation. 
  As can be seen from figure S3 of the SI, for an eKE of \unit[10]{eV} and \unit[20]{eV} a variation of the $\beta$ parameter changes the EMFP little but influences the mean number of elastic scatterings $\langle N_{\rm ela}\rangle = \text{IMFP/EMFP}$ and hence the resulting IMFP.
  The uncertainties given by \citet{thurmer2013} for the $\beta$ parameters of liquid ($\beta_{\rm liq}$) and gas ($\beta_{\rm gas}$) are ca.\ $\pm 0.08$ for an eKE of \unit[20]{eV}.
  According to our sensitivity analysis, the uncertainty of $\beta_{\rm liq}$ leads to an uncertainty in the IMFP of ca.\ \unit[$\pm$20]{\%} (and not by ``a factor of two''; from a simulation, we find that the IMFP changes from \unit[4.6]{nm} to \unit[5.3]{nm}, cf.\ Table \ref{tab:uncert}).
  Similarly, the uncertainty in $\beta_{\rm liq}$ of $\pm 0.15$ at \unit[10]{eV} translates to an uncertainty of \unit[$\pm 60$]{\%} (not ``an order of magnitude''). 
  For completeness, Table \ref{tab:uncert} provides calculated values of EMFP and IMFP for an eKE of \unit[10]{eV} and \unit[20]{eV} when the $\beta$ parameters are changed by the quoted uncertainties.
  These IMFP values are well below those of \citet{michaud2003}, thus inclusion of the uncertainties does not change our conclusions, as expected from our sensitivity analysis.

  \begin{table}
   \begin{tabular}{ c c c | c c } 
            & $\beta_{\rm gas}$ & $\beta_{\rm liq}$ & EMFP (nm) & IMFP (nm) \\ \hline
       10 eV & 0.73             & 0.11              & 0.41      & 5.2 \\
             & 0.73             & 0.27              & 0.66      & 3.2 \\
             & 0.73             & 0.43              & 1.16      & 2.2 \\
             & 0.88             & 0.11              & 0.41      & 5.5 \\
             & 0.88             & 0.27              & 0.56      & 3.8 \\
             & 0.88             & 0.43              & 0.92      & 2.6 \\
             & 1.03             & 0.11              & 0.39      & 6.0 \\
             & 1.03             & 0.27              & 0.52      & 4.3 \\
             & 1.03             & 0.43              & 0.78      & 3.0 
        \\ \hline                                               
       20 eV & 1.36             & 0.38              & 0.77      & 5.0 \\
             & 1.36             & 0.46              & 0.90      & 4.3 \\
             & 1.36             & 0.54              & 1.03      & 3.8 \\
             & 1.44             & 0.38              & 0.73      & 5.3 \\
             & 1.44             & 0.46              & 0.84      & 4.6 \\
             & 1.44             & 0.54              & 0.98      & 4.0 \\
             & 1.52             & 0.38              & 0.68      & 5.8 \\
             & 1.52             & 0.46              & 0.82      & 4.8 \\
             & 1.52             & 0.54              & 0.95      & 4.2
        \\ \hline
    \end{tabular}
    \caption{Determined values for the elastic (EMFP) and inelastic (IMFP) mean free path for different assumed values of the photoelectron angular distribution parameter $\beta$ for ionization of gas-phase water ($\beta_{\rm gas}$) and as measured outside the liquid ($\beta_{\rm liq}$), reflecting the uncertainties reported by \citet{thurmer2013}.}
    \label{tab:uncert}
  \end{table}

  To conclude, we have refuted the four points of the comment. 
  In this reply we have provided additional arguments that support the concepts, methods and results of our work, which have also been documented through references to the relevant contemporary literature. 
  The question why our mean free paths, as well as previous results, such as those discussed by \citet{nikjoo2016,shinotsuka2017}, and \citet{nguyen2018}, are substantially shorter than those reported by \citet{michaud2003} and consequently by \citet{signorell2016} and \citet{luckhaus2017}, is open.
  It may reflect differences between electron scattering in liquid water as compared to amorphous ice, or shortcomings in the employed models or experiments, or a combination of these. 
  In any case, further work is necessary to answer this interesting question with certainty.
  
  \bibliography{bib}{}
  
\end{document}